\documentstyle[12pt,A4]{article}

\begin{document}
\makeatletter
\def\siml{\mathrel{\mathpalette\gl@align<}}
\def\simg{\mathrel{\mathpalette\gl@align>}}
\def\gl@align#1#2{\lower.6ex\vbox{\baselineskip\z@skip\lineskip\z@
 \ialign{$\m@th#1\hfill##\hfil$\crcr#2\crcr{\sim}\crcr}}}
\makeatother
\hbadness=10000
\hbadness=10000
\begin{titlepage}
\nopagebreak
\def\thefootnote{\fnsymbol{footnote}}
\begin{flushright}

        {\normalsize
 Kanazawa-94-10\\
May, 1994   }\\
\end{flushright}
\vspace{2cm}
\vfill
\begin{center}
\renewcommand{\thefootnote}{\fnsymbol{footnote}}
{\large \bf Minimal String Unification and Yukawa Couplings in Orbifold
Models}

\vfill
\vspace{1.1cm}

{\bf Tatsuo Kobayashi
\footnote[1]{e-mail:kobayasi@hep.s.kanazawa-u.ac.jp}
}

\vspace{1.1cm}
       Department of Physics, Kanazawa University, \\

       Kanazawa, 920-11, Japan \\

\vfill
\end{center}
\vspace{1.1cm}

\vfill
\nopagebreak
\begin{abstract}
We study the minimal supersymmetric standard model derived from
$Z_N \times Z_M$ orbifold models.
Moduli dependent threshold corrections of the gauge couplings are
investigated to explain the measured values of the coupling constants.
Also we study Yukawa couplings of the models.
We find that the $Z_2 \times Z_6'$, $Z_2\times Z_6$, $Z_3 \times Z_6$ and
$Z_6 \times Z_6$ orbifold models have the possibility to derive
Yukawa couplings for the second and third generations as well as the measured
gauge coupling constants.
Allowed models are shown explicitly by combinations of modular weights
for the matter fields.

\end{abstract}

\vfill
\end{titlepage}
\pagestyle{plain}
\newpage
\voffset = 0.5 cm

\vspace{0.8 cm}
\leftline{\large \bf 1. Introduction}
\vspace{0.8 cm}

Superstring theory is the best candidate for a unified theory of
all the known interactions including gravity.
All the gauge coupling constants coincide even without a grand unified group
at a string scale $M_{\rm st}=5.27\times g_{\rm st} \times 10^{17}$GeV
\cite{Kaplunovsky}, where $g_{\rm st}\simeq 1/\sqrt 2$ is a universal string
coupling.
Yukawa couplings have the same origin as the gauge coupling constants.

A minimal string model is the 4-dim string vacuum which has the same massless
spectrum as the minimal supersymmetric standard model (MSSM).
That is one of the simplest scenarios to lead from the
string theories to the low energy physics.
The minimal string models do not face the problems of the fast proton decay
and triplet-doublet splitting in the Higgs sector as well as mass splitting
between the quarks and the leptons other than the third generation, unlike
the GUTs or supersymmetric standard models with extra matter fields.
However the minimal string models are not consistent with recent study of the
LEP measurements which show all the gauge coupling constants unify
simultaneously at $M_X\simeq 10^{16}$GeV within the framework of the MSSM
\cite{MSSM}.
We need some threshold corrections at $M_{\rm st}$ in order to explain the
difference between $M_{\rm st}$ and $M_X$.

An orbifold construction is one of the simplest and most interesting methods
to construct 4-dim string vacua \cite{ZNOrbi,ZNM}.
Threshold corrections to gauge couplings of the orbifold models were studied
in ref.\cite{Dixon,Derendinger}.
The correction depends on a moduli parameter $T$, which describes a
geometrical feature of the orbifold like its size.
The parameter is obtained as a vacuum expectation value of a moduli field,
when the supersymmetry (SUSY) breaks.
Refs.\cite{cond,Ross} show that the moduli field take the vacuum expectation
value of order one in the SUSY-breaking due to a gaugino condensation.
Other phenomenological aspects have been studied like Yukawa couplings,
K\"ahler potential and so on.
Further the minimal string models have been searched explicitly
\cite{ZNOrbi2}.

Using the threshold corrections, recent work [10-14] showed the possibility to
derive the gauge coupling constants consistent with the measurements within
the framework of the minimal string models from the orbifold models.
Ref.\cite{Ibanez} showed that $Z_6$-II, $Z_8$-I and $Z_N \times Z_M$ orbifold
models are promising in the case where a level $k_1$ of U(1)$_Y$ is equal
to 5/3.
However string theories obtain any other values of $k_1$ \cite{Ibanez2}.
Refs.\cite{KKO1,KKO2} investigated the case with the general values of $k_1$
and showed the $Z_6$-I orbifold models as well as the above $Z_N$ are
promising, and the $Z_2 \times Z_2$ and $Z_3 \times Z_3$ orbifold models are
ruled out in the case where an overall moduli parameter is of order one and
the SUSY breaks at $M_Z$.
Further it is shown that the $Z_3\times Z_3$ orbifold models are allowed in the
 case where the SUSY-breaking scale is 1TeV.
Thus this constraint to derive the measured gauge coupling constants at $M_Z$
is very useful to select models from the huge number of 4-dim string vacua for
the $Z_N$ orbifold models, while the $Z_N \times Z_M$ orbifold models are not
constrained so much.
Actually explicit information on the allowed $Z_N$ orbifold models is shown
in ref.\cite{KKO2}.

It is very important to extend the above analyses to the prediction of the
Yukawa coupling values for the quarks and the leptons.
Selection rules for the couplings are very restricted in the orbifold models
[16-19].
Thus, before the prediction of the Yukawa coupling values at $M_Z$ it is
useful to study which models allow realistic Yukawa couplings among the
minimal string models obtained at the above stage.
Ref.\cite{KKO2} discussed Yukawa couplings of some models derived from the
$Z_N$ orbifold constructions so as to show the minimal string models can allow
only the Yukawa couplings for the top and bottom quarks as renormalizable
couplings.
The other couplings might be explained due to nonrenormalizable couplings.
Here following the approach of ref.\cite{Casas}, we assume that the second
and third generations of the quarks and the leptons have renormalizable
couplings and the couplings for the first generation could be explained by
nonrenormalizable couplings.
In this paper we study the $Z_N \times Z_M$ models to allow the above type of
the Yukawa couplings among the minimal string models which have the gauge
coupling constants consistent with the measured values at $M_Z$ through
the threshold corrections.
We discuss mainly the cases with $T$ of order one.
Further we show explicitly the models.
That is very useful for model building.

This paper is organized as follows.
In section two we review briefly the $Z_N \times Z_M$ orbifold models.
Then we study which sector are possible to have each MSSM matter field under
a certain value of $k_1$.
Larger values of $k_1$ are required so that there are the matter fields in
the oscillated states.
In section three the threshold corrections of the orbifold models are
reviewed.
We show which modular weights are allowed for the MSSM matter fields under a
 certain value of $k_1$.
In section four we study the possibility to derive minimal string models
consistent with the measurements of the gauge coupling constants.
In section five we investigate the Yukawa couplings allowed in the
minimal string models obtained in section four.
That constrains quite the promising models.
We show explicitly the results.
Section six is devoted to conclusions and discussions.

\vspace{0.8 cm}
\leftline{\large \bf 2. $Z_N\times Z_M$ Orbifold Models}
\vspace{0.8 cm}

In this section we review the $Z_N \times Z_M$ orbifold models \cite{ZNM}.
In the orbifold construction, the string states consist of left-moving and
right-moving bosonic strings on the 4-dim space-time and a 6-dim
$Z_N \times Z_M$ orbifold, a left-moving gauge part and a right-moving
fermionic string which is related to the right-moving bosonic string through
a world-sheet SUSY.
We bosonize the fermionic string so as to obtain the bosonic string whose
momenta span an SO(10) lattice.
Momenta of the gauge parts span an E$_8\times $E$_8'$ lattice.
The $Z_N \times Z_M$ orbifolds are obtained by dividing tori in terms of
two independent twists $\theta$ and $\omega$, where $\theta^N= \omega^M=1$.
We denote eigenvalues of $\theta$ and $\omega$ in a complex basis
($X_i,\tilde X_i$) ($i=1,2,3$) as exp[$2\pi i v^i_1$] and exp[$2\pi i v^i_2$],
 respectively.
The exponents for each $Z_N \times Z_M$ orbifold model are shown in the
second column of Table 1.
When the 6-dim torus is divided by $\theta$ and $\omega$, the SO(10) and
E$_8\times $E$_8'$ lattices are simultaneously divided by some shifts.
Then we obtain only $N=1$ space-time SUSY and a smaller gauge group.
Further the E$_8 \times $E$_8'$ lattice is shifted by Wilson lines
\cite{WL1,WL2,KO3}.
Here we assume that through the above procedure we obtain the
SU(3)$\times$SU(2)$\times$U(1)$_Y$ gauge group in the observable sector.

Closed strings on the orbifolds are classified into two types.
One is an untwisted string and the other is a twisted string.
The former closes even on the torus and has the following massless condition
for the left-mover,
$$ h-1=0,
\eqno(2.1)$$
where $h$ is a conformal dimension due to the E$_8 \times $E$_8'$ gauge part.
The twisted string has a boundary condition twisted by
$\theta^\ell \omega^m$.
Massless states of the $\theta^\ell \omega^m$-twisted sector $T_{\ell m}$
should satisfy the following condition,
$$h+N_{OSC}+c_{\ell m}-1=0,
\eqno(2.2)$$
where $N_{OSC}$ is the oscillator number.
Here $c_{\ell m}$ is the ground state energy obtained as
$$ c_{\ell m}={1 \over 2}\sum^3_{i-1}v^i_{\ell m}(1-v^i_{\ell m}),$$
$$v^i_{\ell m}\equiv \ell v^i_1+mv^i_2-{\rm Int}(\ell v^i_1+mv^i_2),
\eqno(2.3)$$
where Int$(a)$ represents an integer part of $a$.

A state with an $\underline{R}$ representation under a non-abelian group $G$
contributes to the conformal dimension as follows,
$$h={C(\underline {R}) \over C(G)+k},
\eqno(2.4)$$
where $k$ is a level of a Kac-Moody algebra corresponding to $G$ and
$C(\underline{R})$ ($C(G)$) denotes a quadratic Casimir of the
$\underline{R}$ (adjoint) representation.
For example we obtain $C($SU($N))=N$ for the adjoint representation of SU($N$).
The orbifold models in general lead to $k=1$ for the non-abelian group,
although we can obtain the models with higher levels by a complicated
construction \cite{k}.
Therefore we restrict ourselves to the case with $k=1$ for the non-abelian
groups.
On the other hand, the state with a charge $Y$ of the U(1)$_Y$ has another
contribution to the conformal dimension as $Y^2/k_1$, where $k_1$ is the
level for the U(1)$_Y$.
The value of $k_1$ is the free parameter in the minimal string models.
Further the matter fields have charges under extra U(1)s, which might be
broken, and the extra charges contribute to the conformal dimension.

Using the above discussion, we have constraints on massless spectra of the
MSSM matter fields.
For example we consider the quark doublets with $Y=1/6$.
The representation (3,2) under the SU(3)$\times$SU(2) has a contribution
to the conformal dimension by 7/12 through (2.4).
For the quark doublets to have $N_{OSC}$ in $T_{\ell m}$, the level $k_1$
should satisfy the following relation,
$$k_1 \geq {1 \over 36(5/12-c_{\ell m}-N_{OSC})}.
\eqno(2.5)$$
The case with $c_{\ell m}=N_{OSC}=0$ corresponds to the untwisted
sector.
Thus the level $k_1$ should satisfy $k_1>1/15$ so that the quark doublets
appear in the untwisted sector.
Similarly we obtain condition that the quark doublets and the other MSSM
matter fields exist in each sector and have the oscillator number $N_{OSC}$.
Existence of the lepton singlets with $Y=1$ derives the lower bound of $k_1$
 as $k_1\geq 1$.

\vspace{0.8 cm}
\leftline{\large \bf 3. Duality Symmetry and Threshold Corrections}
\vspace{0.8 cm}

Spectra in the orbifold models are invariant under the following
duality transformation \cite{Kikkawa},
$$ T_i \rightarrow {a_iT_i-ib_i \over ic_iT_i+d_i} ,\eqno(3.1)$$
$$a_i,b_i,c_i,d_i\in {\bf Z}, \quad a_id_i-b_ic_i=1,$$
where $T_i$ is a moduli parameter describing a geometrical feature of the
$i$-th plane.
In this paper we restrict ourselves to the case of an overall moduli
parameter, i.e., $T=T_1=T_2=T_3$.

Effective theories derived from the orbifold models also have the duality
symmetry \cite{Ferrara}.
In the theories, the moduli field $T$ have the following K\"ahler potential,
$$ -3{\rm log}|T+\bar T|.
\eqno(3.2)$$
A vacuum expectation value of the moduli field gives the geometry of the
orbifold.
The K\"ahler potential (3.2) is invariant under the duality symmetry (3.1) up
to the K\"ahler transformation.
On the other hand, the K\"ahler potential of the chiral matter field $A$ is
obtained as
$$ (T+\bar T)^{n}A\bar A,
\eqno(3.3)$$
where $n$ is called modular weight \cite{Dixon2,Ibanez}.
The untwisted sector has the modular weight $n=-1$.
The twisted sector with an unrotated plane has the modular weight $n=-1$,
while the twisted sector without unrotated planes has the modular weight
$n=-2$.
Further the oscillator $\partial X_i$ reduces the modular weight by one and
$\partial \tilde X_i$ contributes to the modular weight oppositely.
The duality invariance of (3.3) requires the following duality transformation
of the matter fields,
$$ A \rightarrow A (icT+d)^n.
\eqno(3.4)$$

Using the discussion in the previous section, we have the possible modular
weights for the MSSM matter fields under $k_1$ in each twisted sector of all
orbifold models.
Table 2 and 3 show lower bounds of $k_1$ so that the MSSM matter fields have
each modular weight.
In the tables $Q$, $U$, $D$, $L$, $E$ and $H$ denote the quark doublets,
quark singlets of the up-sector and the down-sector, lepton doublets, lepton
singlets and Higgs fields of the MSSM, respectively.

The duality symmetry becomes anomalous in terms of loop effects due to only
massless fermions \cite{Derendinger2,Derendinger}.
One-loop effective Lagrangian including the duality anomalous term is
obtained as
$$
{\cal L}_{\rm nl}=  \sum_a \int d^2\theta {1 \over 4} W^aW^a[S
 -{1 \over 16\pi^2}{1 \over 16}\Box^{-1}\overline{\cal DD}{\cal DD}
b'_a{\rm log}(T+\overline T)]+{\rm h.c.},
\eqno(3.5)$$
where S is a dilaton/axion field, $W^a$ is a Yang-Milles superfield and $a$
is an index for a gauge group.
The second term of (3.5) is anomalous under the duality symmetry.
Here the duality anomaly coefficients $b'_a$ are obtained as
$$b_a'=-3C(G_a) +\sum_R T(R)(3+n_R),
\eqno(3.6)$$
where $T(R)$ is the Dynkin index for the $R$ representation, i.e., $T(R)=1/2$
for the
$N$-dim fundamental representation of the SU($N$).

The duality anomaly can be cancelled by two ways
\cite{Derendinger2,Derendinger}.
One is the Green-Schwarz (GS) mechanism \cite{GS}, which induces the
non-trivial transformation to the dilaton field $S$ as follows,
$$S \rightarrow S -{1 \over 8 \pi^2} \delta_{GS}{\rm log}
(icT+d),
\eqno(3.7)$$
where $\delta_{GS}$ is an unknown GS coefficient.
It is remarkable that the GS mechanism is independent of the gauge groups.
Further, the duality anomaly can also be cancelled through the moduli
dependent threshold corrections $\Delta (T)$ due to the towers of massive
modes.
The correction is obtained as
$$\Delta_a(T)=-{1 \over 16\pi^2} (b'_a-k_a\delta_{\rm GS})
{\rm log}|\eta(T)|^4,
\eqno(3.8)$$
where $\eta(T)$ is the Dedekind function, i.e.,
$\eta(T)=e^{-\pi T/12}\prod^\infty_{n=1}(1-e^{-2\pi nT}).$

Including the threshold corrections, we have the running gauge coupling
constant $\alpha_a=k_a g_a^2/4 \pi$ at $\mu$ as follows,
$$ \alpha^{-1}_a(\mu)=\alpha^{-1}_{\rm st}+{1 \over 4\pi}{b_a\over k_a}
 {\rm log}{M_{\rm st}^2 \over \mu^2}-{1\over 4\pi}
 ({b'_a \over k_a}-\delta_{\rm GS}){\rm log}[(T+\bar T)|\eta(T)|^4],
 \eqno(3.9)$$
where $\alpha_{\rm st}$ is the universal string coupling and $b_a$ are $N=1$
 $\beta$-function coefficients, i.e., $b_3=-3$, $b_2=1$ and $b_1=11$
for SU(3), SU(2) and U(1)$_Y$ in the MSSM.

We study the unification scale $M_X$ of the SU(3) and SU(2) gauge coupling
constants, $\alpha_3$ and $\alpha_2$.
Note that the gauge coupling constant of U(1)$_Y$, $\alpha_1$ does not always
coincide with the other couplings at $M_Z$, because we consider the case where
the level $k_1$ of U(1)$_Y$ is the general value.
We obtain the unification scale $M_X$ as follows,
$$ {\rm log}(M_X^2/ \mu^2)=\pi \{\sin^2\theta_{\rm W}(\mu)\alpha^{-1}_{\rm em}
 (\mu)-\alpha_3^{-1}(\mu)\}.
\eqno(3.10) $$
Eq.(3.10) is available at $\mu$ higher than the soft SUSY-breaking scale.
We use the measured values of the gauge couplings as
$\sin^2\theta_W(M_Z)=0.2325\pm .0008$,
$\alpha^{-1}_{\rm em}(M_Z)=127.9\pm .1$,
$\alpha_3^{-1}(M_Z)=8.82\pm .27$ at $M_Z=91.173\pm .020$.
If the SUSY breaks at $M_Z$, the gauge couplings of SU(3) and SU(2) coincide
at $M_X=10^{16.2}$GeV.

Using (3.9), we have the following relation between $M_X$ and $M_{\rm st}$
\cite{Ibanez},
$$ {\rm log}{M_{X} \over M_{\rm st}}={1\over 8} \Delta b'
 {\rm log}[(T+\bar T)|\eta(T)|^4] ,
\eqno(3.11)$$
where $\Delta b'\equiv b'_3-b'_2$.
It is remarkable that the value ${\rm log}[(T+\bar T)|\eta(T)|^4]$ is always
negative for any value of $T$.
Therefore we need $\Delta b'>0$ in order to derive $M_X<M_{\rm st}$ from
(3.11).
For example we use $M_X=10^{16.2}$GeV and $M_{\rm st}=3.73\times 10^{17}$GeV
to estimate the value of $T$ in the case with $\Delta b'=3$.
In this case we have $T=11$.

When the SUSY breaks, the moduli field $T$ could take the non-zero
vacuum expectation value.
In the SUSY-breaking scenario due to a gaugino condensation, the value of $T$
has been estimated as $T\sim 1.2$ \cite{cond}.
Further refs.\cite{Ross} take into account a one-loop effective potential to
obtain $T\sim 8$.
Thus it seems that the value of $T$ is of order one.
Therefore we consider mainly the case where $T<11$ and $\Delta b'>3$.

\vspace{0.8 cm}
\leftline{\large \bf 4. Minimal String Model}
\vspace{0.8 cm}

In this section we study the possibility that we derive the minimal string
models consistent with the measured gauge coupling values from the
$Z_N \times Z_M$ orbifold construction.
We assign the allowed modular weights to the MSSM matter fields and
then investigate whether the combinations of the modular weights lead to the
suitable threshold corrections explaining the measured coupling constants.

Using (3.9), we have the following relation between
$\alpha^{-1}_{\rm em} \equiv k_1\alpha^{-1}_1+\alpha^{-1}_2$ and
$\alpha^{-1}_2$,
$$
\begin{array}{cl}
\alpha^{-1}_2(\mu)& ={1\over k_1+1}\alpha^{-1}_{\rm em}(\mu) +
{1 \over 4 \pi} (1 -{12 \over k_1 +1})\log {M_{\rm st}^2 \over \mu^2}\\
& \\
& -{1 \over 4 \pi} (b'_2-{B' \over k_1+1}{\rm log})[(T+\bar T)|\eta(T)|^4],\\
\end{array}
\eqno(4.1)$$
where $B' \equiv b'_1+b'_2$.
{}From (4.1) and (3.11), we derive the following equation of
$\sin^2\theta_W \equiv \alpha_{\rm em}/\alpha_2$,
$$\begin{array}{cl}
\sin ^2\theta_{\rm W}(\mu)& ={1\over k_1+1}+{\alpha_{\rm em}(\mu)\over 4\pi}
\big( (1-{12 \over k_1+1}){\rm log}{M_{\rm string}^2 \over \mu^2}\\
& \\
& -{4\over \Delta b'}(b'_2-{B'\over k_1+1})
{\rm log}{M_X^2 \over M_{\rm string}^2}\big).
\end{array} \eqno(4.2)$$
Then we obtain the following equation,
$$k_1={12\Delta b'{\rm log }(M_{\rm sting}^2 / \mu^2)-4B'{\rm log}
 (M_X^2 / M_{\rm string}^2)- 4 \pi\Delta b'\alpha^{-1}_{\rm em}(\mu) \over
 \Delta b'{\rm log }(M_{\rm sting}^2 / \mu^2)-4b'_2{\rm log}
 (M_X^2 / M_{\rm string}^2)-4 \pi\Delta b'\alpha^{-1}_{\rm em}(\mu)
 {\rm sin}^2\theta_W(\mu)}-1,
\eqno{(4.3)} $$
where we obtain the unification scale $M_X$ from (3.10) for the case of any
SUSY-breaking scale.

At first we assign the allowed modular weights to all the MSSM matter
fields, i.e., $Q$, $U$, $D$, $L$, $E$ and $H$.
We calculate the duality anomaly coefficients of their combinations and
restrict ourselves to the case with $\Delta b' >3$.
Then for each combination we estimate the value of $k_1$, using (4.3) and
(3.10).
Here we have to investigate whether or not each combination includes
modular weights allowed by this level $k_1$ through Table 2 and Table 3.

We estimate the allowed values of $k_1$ in the cases where the SUSY breaks
at $M_Z$ and 1TeV.
In the third column of Table 1, $M_S$ represents the SUSY-breaking scale.
In the case with the SUSY-breaking at 1TeV, we use
$\alpha_{\rm em}^{-1}({\rm 1\ TeV}) = 127.2\pm 0.1$,
$\sin^2\theta _{\rm W}({\rm 1\ TeV})=0.2432\pm 0.0021$ and
 $\alpha^{-1}_3({\rm 1\ TeV})=11.48\pm 0.27$,  which are calculated from the
values at $M_Z$ through the renormalization group equation in the
non-supersymmetric standard model, i.e., $b_3=-7$, $b_2=-19/6$ and
$b_1=41/6$.
The results are found in Table 1, whose fourth column
shows the maximum values of $\Delta b'$ and the corresponding values of $T$
 in the allowed models and fifth column shows the allowed ranges of $k_1$
 \footnote{
The ranges of $k_1$ obtained in ref.\cite{KKO1} include a minor mistake, and
should be replaced by the results shown in Table 1.}.
We cannot derive consistent levels from the $Z_3 \times Z_3$ orbifold models
 in the case where $\Delta b'>3$ and the SUSY breaks at $M_Z$.
We need at most $\Delta b'=2$ to derive the measured gauge coupling constants.
The range of $k_1$ in the parenthesis of the fifth column represents the
corresponding values of $k_1$ in the case with $\Delta b'=2$.

The $Z_2 \times Z_2$ orbifold models are ruled out, because they always
derive $\Delta b'<0$.
At this stage, all the orbifold models except the $Z_2 \times Z_2$ are
promising for the minimal string model consistent
with the measured gauge coupling constants.
We find the range of $k_1$ as $1 \siml k_1 \siml 2$.
Under these values of $k_1$, some of oscillated states are ruled out.
The above estimation of the levels includes at most 20\% experimental error.
Some of the models are possible to have $k_1=5/3$, which is the level
predicted by GUTs.
It is remarkable that the minimal string models from the $Z_N \times Z_M$
orbifold constructions can obtain smaller value of $T$ than ones from the
$Z_N$ orbifolds.
The latter has the duality anomaly cancellation condition, which constrains
the massless spectra.

\vspace{0.8 cm}
\leftline{\large \bf 5. Yukawa Coupling}
\vspace{0.8 cm}

It is an important problem to derive the realistic Yukawa couplings in
addition to the gauge coupling constants at $M_Z$.
The orbifold models have selection rules for the Yukawa couplings [16-19].
A point group selection rule requires that a product of point group elements
should be an identity.
Moreover, allowed couplings should conserve the SO(10) momenta of the
right-moving bosonized fermionic string.
Sectors allowed to couple are shown explicitly in ref.\cite{KO2}.
Further the $Z_N$ invariance requires a product of the $Z_N$ phases from the
oscillated states to be zero for each plane.
Thus the Yukawa couplings are very restricted in the orbifold models,
especially for the oscillated states.
Actually the minimal string models derived from the $Z_N$ orbifold models
in ref.\cite{KKO2} allow at most the Yukawa couplings for the top and bottom
quarks as renormalizable couplings.
Also, some Yukawa couplings at the low energy could be obtained by
nonrenormalizable couplings.
However they are suppressed by $1/M_{\rm pl}$.
Therefore we should explain the large values of the Yukawa couplings
by the renormalizable couplings.
Following ref.\cite{Casas}, we assume that the Yukawa couplings for the
second and third generations are due to the renormalizable couplings.
Then we search the minimal string models to allow these types of couplings
among the combinations of the modular weights obtained at the previous stage.

For example we study the Yukawa couplings of the minimal string models
derived from the $Z_2 \times Z_4$ orbifold construction.
The point group selection rule and the SO(10) momentum conservation forgive
the following couplings \cite{KO2},
$$U_1U_2U_3, \quad U_1T_{01}T_{03}, \quad U_1T_{02}T_{02}, \quad
U_1T_{10}T_{10},$$
$$U_3T_{12}T_{12}, \quad T_{02}T_{10}T_{12},
\eqno(5.1)$$
where $U_i$ denotes the untwisted sector associated with the $i$-th plane.
Eq.(5.1) corresponds to the couplings where all of the states have $n=-1$.
Also the $Z_2 \times Z_4$ orbifold models allow the following couplings,
$$T_{01}T_{11}T_{12}, \quad T_{03}T_{10}T_{11}, \quad T_{11}T_{11}T_{02}.
\eqno(5.2)$$
The first and second types of (5.2) correspond to the couplings of states
with $n=-1$, --1 and --2, while the last corresponds to the coupling of
states with $n=-1$, --2 and --2.

Further we have to consider the couplings of the oscillated states.
The $Z_2 \times Z_4$ orbifold models do not allow the quark doublets to
have the non-zero oscillator number, although other matter fields can be
obtained from the oscillated states.
For each plane, a product of the $Z_N$ phases due to the oscillators should
vanish in order to allow the couplings.
We study this selection rule in the $Z_2 \times Z_4$ orbifold models to find
that the quark singlets and the Higgs fields in the oscillated states are
not allowed to couple with the quark doublets.
Similarly the leptons in the oscillated states are impossible to couple with
the Higgs fields which have the vanishing oscillator number.
Namely any oscillated state does not have couplings.
As the results, the allowed couplings are represented by the modular weights
 as follows,
$$(n_1,n_2,n_3)=(-1,-1,-1), \quad (-1,-1,-2), \quad (-1,-2,-2).
\eqno(5.3)$$

Similarly we investigate the combinations of the modular weights which allow
the Yukawa couplings for the other orbifold models.
Some orbifold models allow the quark doublets as well as other matter fields
in the oscillated states.
However, the couplings of the oscillated states are impossible in the case
 with $1\leq k_1 \leq 2.06$, because the $Z_N$ invariance in general requires
the larger oscillation number for the oscillated states to couple and the
values of $k_1$ constrain the larger oscillator number of the states.
The $Z_2 \times Z_6$ orbifold models have the same combinations of the
modular weights to couple as the $Z_2 \times Z_4$, i.e., (5.3).
The $Z_4 \times Z_4$, $Z_3 \times Z_6$ and $Z_6 \times Z_6$ orbifold models
allow the type of the couplings with $(n_1,n_2,n_3)=(-2,-2,-2)$ in addition to
(5.3).
The $Z_3 \times Z_3$ orbifold models have the couplings as,
$$(n_1,n_2,n_3)=(-1,-1,-1), \quad (-1,-1,-2), \quad (-2,-2,-2),
\eqno(5.4)$$
while the $Z_2 \times Z_6'$ orbifold models allow the following couplings,
$$(n_1,n_2,n_3)=(-1,-1,-1), \quad (-1,-2,-2), \quad (-2,-2,-2).
\eqno(5.5)$$
Note that the $Z_6 \times Z_6$ have the same constraints
on the modular weights as the $Z_3 \times Z_6$, and both allow the same types
of the Yukawa couplings.

Here we study the minimal string models to allow the Yukawa couplings.
For example we study the minimal string models derived from the
$Z_2 \times Z_4$ orbifold construction in the case with the SUSY-breaking
at $M_Z$ and $\Delta b'>3$.
At the previous stage, these models allow $1.00\leq k_1 \leq 1.60$ and
$\Delta b'\leq 12$, which are shown in the third column of Table 4 as well as
 Table 1.
If we require the top Yukawa coupling as the renormalizable coupling,
the minimal string models are constrained and we obtain
$1.00 \leq k_1 \leq 1.58$
and $\Delta b'\leq 10$, which are shown in the fourth column of Table 4.
Further the Yukawa couplings for the bottom quark as well as the top are
allowed in the models with $1.03 \leq k_1 \leq 1.54$ and $\Delta b'\leq 9$,
which are shown in the fifth column of Table 4.
Similarly the sixth column of the table shows the models to allow
the Yukawa couplings for the charm quark as well as the third generation of
the quarks, and the seventh column lists the models with the Yukawa couplings
for the second and third generations of the quarks.
Moreover, the eighth and ninth columns show the models to forgive the Yukawa
couplings for the one and two leptons, respectively, in addition to the
Yukawa couplings for the two generations of the quarks.
The lepton couplings are impossible for the minimal string models derived
from  the $Z_2 \times Z_4$ orbifold models in the case where the SUSY breaks
at $M_Z$ and $\Delta b'>3$.

Similarly we obtain the minimal string models to allow the Yukawa couplings
from the other orbifold models in the case with the SUSY-breaking at $M_Z$
and $\Delta b'>3$.
The results are shown in Table 4.
Further Table 5 shows the results in the case with SUSY-breaking at 1TeV and
$\Delta b'>3$.
These final results are also shown in the sixth and seventh columns of
Table 1.
If we forgive the case with $0 < \Delta b' \leq 3$, we find the minimal
string models to allow the Yukawa couplings for the two generations of the
 quarks and the leptons among the $Z_2 \times Z_4$ and $Z_4 \times Z_4$
orbifold models.
The Yukawa couplings for the two generations are allowed in the
$Z_2 \times Z_4$ orbifold models with $\Delta b'=1$ and the SUSY-breaking at
1TeV.
The range of $k_1$ for this case is shown in the parenthesis of the seventh
column of Table 1.
The Yukawa couplings for the two generations are possible
in the $Z_4 \times Z_4$ orbifold models with at most $\Delta b'=2$.
For $\Delta b'=2$, the corresponding values of $k_1$ are shown in the
parentheses of the seventh column.
In the case with the SUSY-breaking at $M_Z$ there are four types of the
combinations in the $Z_4 \times Z_4$ orbifold models.
In all the types, every modular weight of $E$ is equal to $-2$, and four
modular weights among $L$ and $H$ are assigned as $n=-2$ and the other has
$n=-3$.
The modular weights for the other matter fields are assigned as follows,
$$-1,-1,-2 {\rm \ for \ }Q, \quad -1,-2,-2 {\rm \ for \ }U, \quad
0,-1,-1 {\rm \ for \ }D{\rm \ in \ Type \ 1},$$
$$-1,-2,-2 {\rm \ for \ }Q, \quad -2,-2,-2 {\rm \ for \ }U, \quad
0,-1,-1 {\rm \ for \ }D{\rm \ in \ Type \ 2},$$
$$-2,-2,-2 {\rm \ for \ }Q, \quad -2,-2,-2 {\rm \ for \ }U, \quad
-1,-1,-1 {\rm \ for \ }D{\rm \ in \ Type \ 3},$$
$$-2,-2,-2 {\rm \ for \ }Q, \quad -2,-2,-2 {\rm \ for \ }U, \quad
0,-1,-2 {\rm \ for \ }D{\rm \ in \ Type \ 4}.$$
It is remarkable that Type 3 includes only non-oscillated states in the
quark sector.

We find that the $Z_2 \times Z_6'$, $Z_2 \times Z_6$, $Z_3 \times Z_6$ and
$Z_6 \times Z_6$ orbifold models are promising for the minimal string
model to derive the realistic Yukawa couplings.
These models derive $1.00 \leq k_1 \leq 1.51$, $\Delta b' \leq 8$ and
$T\geq 5.3$.
The values of $k_1$ shown in the last columns of Table 4 and 5 include
the 15\%, 10\% and 13\% experimental errors for the $Z_2 \times Z_6'$,
$Z_2 \times Z_6$ and $Z_3 \times Z_6$ orbifold models, respectively.
The values of $k_1$ and $\Delta b'$ as well as the combinations of the
modular weights are constrained much more than the cases without the
restrictions on the Yukawa couplings.
The case with $k_1=5/3$ is ruled out for any orbifold model.

To show explicitly the obtained models is very useful for model building.
In the case of the $Z_2 \times Z_6'$ ($Z_2 \times Z_6$) orbifold models with
$\Delta b' >3$, Table 6 (7) shows the combinations of the modular weights for
the minimal string models which derive the measured gauge coupling constants
 with the SUSY-breaking at $M_Z$ and allow the Yukawa couplings for the two
generations of the quarks and the leptons.
Note that in Table 6 the combinations \# 2, 4 and 5 do not include the
oscillated states in the quark sector.
Similarly Table 8-1 shows the combinations of the modular weights for the
minimal string models derived from the $Z_3 \times Z_6$ and $Z_6 \times Z_6$
orbifold models with $\Delta b' \geq 6$ and the SUSY-breaking at $M_Z$.
It is found in the table that we can obtain the allowed combinations
of the modular weights with the values of $\Delta b'$ decreasing by one from
the allowed combinations with the larger values of $\Delta b'$ through the
following recipe of the substitutions,
$$\begin{array}{ccrrl}
 (1)& \quad  n& = &-2 \rightarrow & n=-1 \quad {\rm in \ }Q,\\
 (2)& \quad  n& = &0 \rightarrow & n=-1 \quad {\rm in \ }U,\\
 (3)& \quad  n& = & -1 \rightarrow & n=-2 \quad {\rm in \ }U,
\end{array}
$$
where (1) is possible except the combinations with $n=-1$ for all of the three
lepton singlets.
For example the model of \#2, 7 and 10 in Table 8-1 are obtained from \#1 with
$\Delta b'=8$ through the substitutions (1), (2) and (3), respectively.
The minimal string models with $\Delta b'=5$ and 4 are shown in Table 8-2 and
Table 8-3, where we omit the combinations obtained from the models with
$\Delta b'=6$ through the above substitutions.
In Table 8-2 (8-3), the 32 (69) combinations with $\Delta b'=5$ (4) are
omitted.
In Table 8-2, the combination \# 6 with $\Delta b'=5$ corresponds to the model
where all of the quarks have the vanishing oscillator number.
We obtain the minimal string model from \#6, through the substitution (3),
although the obtained model with $\Delta b'=4$ is omitted in Table 8-3.
In this model all of the quarks belong to the non-oscillated states.

At last we comment on the mass hierarchy of the quarks and leptons.
In the orbifold models, renormalizable couplings of the twisted states are
often suppressed by a contribution due to a world-sheet instanton as
$e^{-aT}$, where $a$ dependens on a distance between fixed points of the
states to couple \cite{Yukawa,KO3}.
This property could explain the mass hierarchy \cite{mass,Casas}.
For example the $Z_2 \times Z_6'$ orbifold models allow the following
couplings \cite{KO2},
$$ U_1U_2U_3, \quad U_1T_{13}Y_{13}, \quad U_2T_{10}Y_{10}, \quad
U_3T_{03}Y_{03}, \quad T_{03}T_{10}Y_{13},
\eqno(5.6)$$
as well as
$$T_{01}T_{02}Y_{03}, \quad T_{02}T_{10}Y_{14}, \quad T_{02}T_{11}Y_{13},
\quad T_{01}T_{11}Y_{14}, \quad T_{02}T_{02}Y_{02}.
\eqno(5.7)$$
Eq.(5.6) corresponds to the couplings of the states with
$(n_1,n_2,n_3)=(-1,-1,-1)$ and these couplings do not have the suppression
factor due to the world-sheet instanton.
Therefore these couplings are not available for the Yukawa couplings for the
second generation, although the other couplings (5.7) could be suppressed
and possible to explain the suppressed Yukawa couplings for the second
generation.
Thus in realistic models the Yukawa couplings for the second generation have
to belong to the types of (5.7).
That might give another constraint on the minimal string model.
However we can find the couplings of (5.7) for the second generation in any
combination obtained in Table 6.
Similarly the models obtained from the $Z_2 \times Z_6$, $Z_3 \times Z_6$ and
$Z_6 \times Z_6$ orbifold constructions could always derive the suppressed
Yukawa couplings for the second generation.
It is very interesting to study assignments of fixed points to the MSSM matter
 fields to explain the mass hierarchy among the minimal string models obtained
the above, as discussed in ref.\cite{Casas}.
That will be investigated elsewhere.

\vspace{0.8 cm}
\leftline{\large \bf 6. Conclusions and Discussions}
\vspace{0.8 cm}

We have studied the minimal string models which have the gauge coupling
constants consistent with the measurements at $M_Z$.
The Yukawa couplings allowed in these models are also investigated.
The $Z_2 \times Z_6'$, $Z_2 \times Z_6$, $Z_3 \times Z_6$ and
$Z_6 \times Z_6$ orbifold models are promising for the minimal string models
 with the Yukawa couplings of the two generations as the renormalizable
couplings.
These models derive $1.00 \leq k_1 \leq 1.51$, $\Delta b' \leq 8$ and
$T \leq 5.3$.
Tables 6, 7, 8-1, 8-2 and 8-3 show the combinations of the modular weights
for the allowed models, explicitly.
That is useful for model building.
Especially it is very interesting to study which assignments of the MSSM
matter fields to the fixed points derive the realistic mass hierarchy of the
quarks and the leptons among the models shown in the tables.

In the above analysis, we have assumed that all the soft masses are equal to
$M_Z$ or 1TeV.
However the string theories in general derive the non-universal soft
SUSY-breaking terms \cite{non}.
Ref.\cite{KSY} shows that the unification
scale $M_X$ of the SU(3) and SU(2) gauge coupling constants depends on the
non-universality of the soft SUSY-breaking terms.
The unification scale of the non-universal case is often higher than
one of the universal case.
In some cases with the non-universal soft masses, the values of
$\Delta b'=2$ or 1 could lead to $T$ of order one.
Thus it is important to study the minimal string model taking into account
the non-universality.
Further it is interesting to extend to the case where $T_i$ have different
vacuum expectation values one another.
In a similar way we can investigate the extended MSSM with extra matter
fields like ref.\cite{extra}.

\vspace{0.8 cm}
\leftline{\large \bf Acknowledgement}
\vspace{0.8 cm}

The author would like to thank H.~Kawabe, N.~Ohtsubo, D.~Suematsu and
Y.~Yamagishi for useful discussions.
This work is partially supported by Soryuushi Shogakukai.



\newpage
\pagestyle{empty}
\noindent
{\bf \large Table 1. Restricted values of $k_1$ and $T$}

\vspace{5mm}

\begin{tabular}{|c|c||c|c|c|c|c|}
\hline
Orbifold           & $v_1$      & $M_S$      & \multicolumn{2}{c|}{Gauge}
& \multicolumn{2}{c|} {Yukawa} \\ \cline{4-7}
 & $v_2$ & & $\Delta b'$  ($T$) & $k_1$   &  $\Delta b'$ ($T$) & $k_1$ \\
\hline \hline
$Z_2\times Z_2$
  & (1,0,1)/2  &
  $M_Z$ & ---  &   --- &   ---   &  ---   \\
  & (0,1,1)/2  &
  1TeV & ---   &   --- &   ---   &    ---     \\ \hline
$Z_3\times Z_3$
  & (1,0,2)/3 &
  $M_Z$ &  2 (15)  & (1.16-1.53)  & ---  &   ---      \\
  & (0,1,2)/3 &
  1TeV& 7 (5.8) &  1.07-1.35  & ---  &   ---      \\ \hline
$Z_2\times Z_4$
  & (1,0,1)/2 &
  $M_Z$ & 12 (4.0) & 1.00-1.60 & --- & ---  \\
  & (0,1,3)/4 &
  1TeV&  12 (4.0)   & 1.00-1.62 & 1 (28) &  (1.00--1.49)    \\ \hline
$Z_4\times Z_4$
  & (1,0,3)/4  &
  $M_Z$ & 12 (4.0) & 1.00-1.60 & 2 (15) & (1.00-1.51) \\
  & (0,1,3)/4  &
  1TeV& 12 (4.0)   & 1.00-1.62   & 2 (15) & (1.12--1.51)  \\ \hline
$Z_2\times {Z_6}'$
  & (1,0,1)/2  &
  $M_Z$ & 12 (4.0) & 1.00-1.92  & 6 (6.4)  & 1.00--1.39 \\
  &(1,1,4)/6  &
  1TeV & 12 (4.0) &  1.00-1.91  & 6 (6.4) &  1.00--1.42  \\ \hline
$Z_2\times Z_6$
  & (1,0,1)/2  &
  $M_Z$ & 18 (3.1) & 1.00-1.71  & 7 (5.8)  & 1.08--1.31 \\
  & (0,1,5)/6 &
  1TeV& 18 (3.1) &   1.00-1.71  & 8 (5.3) &   1.02--1.34     \\ \hline
$Z_3\times Z_6$
  & (1,0,2)/3  &
  $M_Z$ & 18 (3.1) & 1.00-2.06 & 8 (5.3) & 1.01--1.45 \\
  &(0,1,5)/6 &
  1TeV & 18 (3.1) &  1.00-2.04  & 8 (5.3) &   1.02--1.51   \\ \hline
$Z_6\times Z_6$  &(1,0,5)/6 & \multicolumn{5}{c|}{same as $Z_3\times Z_6$}\\
& (0,1,5)/6 & \multicolumn{5}{c|}{ }         \\ \hline
                                   \end{tabular}

\vspace{10mm}

{\bf \large Table 2. Lower-bound of $k_1$ in twisted sectors (I)}
\vspace{5mm}

\begin{tabular}{|c|c|c||r|c|c|c|c|c|c|}
\hline
 \multicolumn{3}{|c||}{T-sec.}
        &     & \multicolumn{5}{c|}{Lower-bound of $k_1$} \\ \hline
 $Z_2\times Z_2$  & $Z_2\times Z_4$ &  $Z_4\times Z_4$
        & $n$ & $Q$ & $D$ & $U$ & $L,H$ & $E$  \\ \hline \hline
         -        & $T_{01},T_{03}$ & $T_{01}$,$T_{03}$,
        &$-1$ &  4/33   &    16/69      &    64/69      &    4/9  & 16/13 \\
                  &                 & $T_{10}$,$T_{13}$,
        &$-2$,0 &    -    &    16/33      &    64/33      &    4/5  & 16/9  \\
                  &                 & $T_{30}$,$T_{31}$
        &$-3$,1 &    -    &       -       &      -        &    4    & 16/5  \\
                  &                 &
        &$-4$,2 &    -    &       -       &      -        &    -    & 16    \\
\hline
$T_{01},T_{10},$  & $T_{02},T_{10}$,& $T_{02}$,$T_{20}$,
        &$-1$ &  1/6    &    4/15       &    16/15      &    1/2  & 4/3   \\
$T_{11}$          & $T_{12}$        & $T_{22}$
        &$-2$,0 &    -    &       -       &      -        &    -    & 4     \\
\hline
     -            & $T_{11}$        & $T_{11}$,$T_{12}$,
        &$-2$ &  4/15   &   16/51       &    64/51      &    4/7  & 16/11 \\
                  &                 &  $T_{21}$
        &$-3$ &    -    &   16/15       &    64/15      &    4/3  & 16/7  \\
                  &                 &
        &$-1$,$-4$ &    -    &       -       &      -     &    -    & 16/3  \\
\hline
                                   \end{tabular}

\newpage

{\bf \large Table 3. Lower-bound of $k_1$ in twisted sectors (II)}

\vspace{5mm}
\footnotesize
\begin{tabular}{|c|c|c|c|c||c|c|c|c|c|c|c|}
\hline
 \multicolumn{5}{|c||}{T-Sec.} &     & \multicolumn{5}{c|}{Lower-bound of
 $k_1$}  \\ \hline
 $Z_3\times Z_3$  & $Z_2\times {Z_6}'$ &  $Z_2\times Z_6$ &  $Z_3\times Z_6$
 &  $Z_6\times Z_6$
  & $n$ & $Q$ & $D$ & $U$ & $L,H$ &
 $E$  \\ \hline \hline
        -         &        -           & $T_{01},T_{05}$  &
 $T_{01},T_{05}$  &  $T_{01}$,$T_{05}$,
  &$-1$ &  1/10   &     4/19      &    16/19      &    9/22 & 36/31 \\
                  &                    &                  &
                  &  $T_{10}$,$T_{15}$,
  &$-2$,0 &   1/4   &     4/13      &    16/13      &    9/16 & 36/25 \\
                  &                    &                  &
                  & $T_{50}$,$T_{51}$
  &$-3$,1 &    -    &     4/7       &    16/7       &    9/10 & 36/19 \\
                  &                    &                  &
                  &
  &$-4$,2 &    -    &      4        &      16       &    9/4  & 36/13 \\
                  &                    &                  &
                  &
  &$-5$,3 &    -    &       -       &      -        &   -     & 36/7  \\
                  &                    &                  &
                  &
  &$-6$,4 &    -    &       -       &      -        &    -    & 36   \\ \hline
$T_{01},T_{02}$,  &         -          & $T_{02},T_{04}$  &
$T_{02},T_{04}$,  & $T_{02}$,$T_{04}$,
  &$-1$ &  1/7    &    1/4        &      1        &    9/19 & 9/7   \\
$T_{10},T_{12}$,  &                    &                  &
$T_{10},T_{14}$,  & $T_{20}$,$T_{24}$,
  &$-2$,0 &    -    &      1        &      4        &    9/7  & 9/4   \\
$T_{20},T_{21}$   &                    &                  &
$T_{20},T_{22}$   & $T_{40}$,$T_{42}$
  &$-3$,1 &    -    &       -       &      -        &    -    & 9   \\ \hline
       -          & $T_{03},T_{10}$,   & $T_{03},T_{10}$, &
$T_{03}$          & $T_{03}$,$T_{30}$,
  &$-1$ &  1/6    &    4/15       &    16/15      &    1/2  & 4/3   \\
                  & $T_{13}$           & $T_{13}$         &
                  &    $T_{33}$
  &$-2$,0 &    -    &       -       &      -        &    -    & 4   \\ \hline
       -          &          -         & $T_{11},T_{12}$, &
$T_{11},T_{13}$   & $T_{12}$,$T_{13}$,
  &$-2$ &  1/4    &    4/13       &     16/13     &    9/16 & 36/25   \\
                  &                    &                  &
                  & $T_{21}$,$T_{23}$,
  &$-3$ &    -    &    4/7        &     16/7      &    9/10 & 36/19   \\
                  &                    &                  &
                  & $T_{31}$,$T_{32}$
  &$-4$ &    -    &      4        &      16       &    9/4  & 36/13   \\
                  &                    &                  &
                  &
  &$-1$,$-5$ &    -    &       -       &      -        &    -    & 36/7    \\
                  &                    &                  &
                  &
  &$-6$ &    -    &       -       &      -        &    -    & 36     \\ \hline
 $T_{11}$         & $T_{02}$           &         -        &
 $T_{12}$         & $T_{22}$
  &$-2$ &  1/3    &    1/3        &     4/3       &    3/5  & 3/2   \\
                  &                    &                  &
                  &
  &$-3$ &    -    &       -       &      -        &    3    & 3     \\ \hline
       -          & $T_{01},T_{11}$,   &         -        &
 $T_{21}$         & $T_{11}$,$T_{14}$,
  &$-2$ &  1/6    &     4/15      &    16/15      &    1/2  & 4/3    \\
                  & $T_{14}$           &                  &
                  & $T_{41}$
  &$-3$ &    -    &     4/9       &    16/9       &    3/4  & 12/7   \\
                  &                    &                  &
                  &
  &$-1$,$-4$ &    -    &     4/3       &    16/3       &    3/2  & 12/5   \\
                  &                    &                  &
                  &
  &$-5$ &    -    &       -       &      -        &   -     & 4      \\
                  &                    &                  &
                  &
  &$-6$,0 &    -    &       -       &      -        &    -    & 12   \\ \hline
                                   \end{tabular}


\newpage

{\bf \large Table 4. Values of $k_1$ and $\Delta b'$ with SUSY-breaking at
$M_Z$}

\vspace{5mm}
\footnotesize
\begin{tabular}{|c|c|c|c|c|c|c|c|c|}
\hline
Orbifold & & \multicolumn{7}{c|}{ Coupling} \\ \cline{3-9}
& & & t & b & c & s & $\tau$ & $\mu$ \\ \hline \hline
$Z_2 \times Z_4$ & $\Delta b'$ & 12 (4.0) & 10 (4.5) &  9 (4.9) &  9 (4.9)
&  7 (5.8) & --- & --- \\
&$k_1$& 1.00-1.60 & 1.00-1.58 & 1.03-1.54 & 1.03-1.54 & 1.03-1.51 &
--- & --- \\ \hline
$Z_4 \times Z_4$ & $\Delta b'$ & 12 (4.0) &11 (4.2) &  9 (4.9) &  9 (4.9)
&  7 (5.8) & --- & --- \\
&$k_1$& 1.00-1.60 & 1.00-1.58 & 1.00-1.54 & 1.03-1.54 & 1.03-1.51 &
--- & --- \\ \hline
$Z_2 \times Z_6'$ & $\Delta b'$ & 12 (4.0) & 10 (4.5) &  9 (4.9) &  8 (5.3)
&  8 (5.3) & 7 (5.8) & 6 (6.5) \\
&$k_1$& 1.00-1.92 & 1.00-1.85 & 1.00-1.59 & 1.00-1.59 & 1.00-1.59 &
1.00-1.53 & 1.00-1.39 \\ \hline
$Z_2 \times Z_6$ & $\Delta b'$ & 18 (3.1) & 16 (3.3) & 13 (3.8) & 12 (4.0)
& 10 (4.5) & 9 (4.9) & 7 (5.8) \\
&$k_1$& 1.00-1.71 & 1.00-1.65 & 1.00-1.59 & 1.00-1.59 & 1.00-1.54 &
1.01-1.35 & 1.08-1.31 \\ \hline
$Z_3 \times Z_6$ & $\Delta b'$ & 18 (3.1) & 16 (3.3) & 13 (3.8) & 12 (4.0)
& 10 (4.5) & 9 (4.9) & 8 (5.3) \\
($Z_6 \times Z_6$)&$k_1$& 1.00-2.06 & 1.00-1.94 & 1.00-1.91 & 1.00-1.82
& 1.00-1.82 & 1.01-1.58 & 1.01-1.45 \\ \hline
\end{tabular}
\vspace{10mm}

{\bf \large Table 5. Values of $k_1$ and $\Delta b'$ with SUSY-breaking at
1TeV}

\vspace{5mm}
\footnotesize
\begin{tabular}{|c|c|c|c|c|c|c|c|c|}
\hline
Orbifold & & \multicolumn{7}{c|}{ Coupling} \\ \cline{3-9}
& & & t & b & c & s & $\tau$ & $\mu$ \\ \hline \hline
$Z_3 \times Z_3$ & $\Delta b'$ & 7 (5.8) & 6 (6.5) & --- & --- & --- & ---
& --- \\
&$k_1$& 1.07-1.35 & 1.07-1.35 & --- & --- & --- &
--- & --- \\ \hline
$Z_2 \times Z_4$ & $\Delta b'$ & 12 (4.0) & 11 (4.2) &  9 (4.9) &  9 (4.9)
&  8 (5.3) &  7 (5.8)  & --- \\
&$k_1$& 1.00-1.62 & 1.00-1.60 & 1.03-1.56 & 1.03-1.56 & 1.03-1.52 &
1.09-1.36 & --- \\ \hline
$Z_4 \times Z_4$ & $\Delta b'$ & 12 (4.0) & 11 (4.2) &  9 (4.9) &  9 (4.9)
&  8 (5.3) & 7 (5.8) & --- \\
&$k_1$& 1.00-1.62 & 1.00-1.60 & 1.03-1.56 & 1.03-1.56 & 1.03-1.53 &
1.03-1.49 & --- \\ \hline
$Z_2 \times Z_6'$ & $\Delta b'$ & 12 (4.0) & 10 (4.5) &  9 (4.9) &  8 (5.3)
&  8 (5.3) & 7 (5.8) & 6 (6.5) \\
&$k_1$& 1.00-1.91 & 1.00-1.85 & 1.00-1.72 & 1.00-1.72 & 1.00-1.72 &
1.00-1.55 & 1.00-1.42 \\ \hline
$Z_2 \times Z_6$ & $\Delta b'$ & 18 (3.1) & 16 (3.3) & 13 (3.8) & 12 (4.0) &
 10 (4.5) & 9 (4.9) & 8 (5.3) \\
&$k_1$& 1.00-1.71 & 1.00-1.66 & 1.00-1.61 & 1.00-1.61 & 1.00-1.56 &
1.01-1.46 & 1.02-1.34 \\ \hline
$Z_3 \times Z_6$ & $\Delta b'$ & 18 (3.1) & 16 (3.3) & 13 (3.8) & 12 (4.0)
& 10 (4.5) & 9 (4.9) & 8 (5.3) \\
($Z_6 \times Z_6$)&$k_1$& 1.00-2.04 & 1.00-1.93 & 1.00-1.91 & 1.00-1.82
& 1.00-1.82 & 1.01-1.60 & 1.02-1.51 \\ \hline
\end{tabular}

\newpage
\normalsize
{\bf \large Table 6. Minimal String Models from $Z_2 \times Z_6'$ orbifold}
\vspace{5mm}

\begin{tabular}{|c|c|c|c|c|c|c|}
\hline
 \#  & $\Delta b'^k$  & Q  & U & D & L, H & E \\  \hline \hline
  1  &  6   & $-2,-2,-3$ & $-1,-1,-2$ & $-1,-1,-1$
& $-2,-2,-2-,2,-3$ & $-2,-2,-2$ \\ \hline
  2  &  5   & $-2,-2,-2$ & $-1,-1,-1$ & $-1,-1,-1$
& $-2,-2,-2-,2,-3$ & $-2,-2,-2$ \\
  3  &  5   & $-2,-2,-3$ & $-1,-1,-2$ & $-1,-1,-1$
& $-2,-2,-2-,2,-3$ & $-2,-2,-2$ \\ \hline
  4  &  4   & $-1,-2,-2$ & $-1,-1,-1$ & $-1,-1,-1$
& $-2,-2,-2-,2,-3$ & $-2,-2,-2$ \\
  5  &  4   & $-2,-2,-2$ & $-1,-1,-2$ & $-1,-1,-1$
& $-2,-2,-2-,2,-3$ & $-2,-2,-2$ \\
  6  &  4   & $-2,-2,-3$ & $-1,-1,-2$ & $-1,-1,-2$
& $-2,-2,-2-,2,-3$ & $-2,-2,-2$ \\
  7  &  4   & $-2,-2,-3$ & $-1,-2,-2$ & $-1,-1,-1$
& $-2,-2,-2-,2,-3$ & $-1,-2,-2$ \\
  8  &  4   & $-2,-2,-3$ & $-1,-2,-2$ & $-1,-1,-1$
& $-2,-2,-2-,2,-3$ & $-2,-2,-2$ \\
\hline
                                   \end{tabular}

\vspace{10mm}
{\bf \large Table 7. Minimal String Models from $Z_2 \times Z_6$ orbifold}
\vspace{5mm}

\begin{tabular}{|c|c|c|c|c|c|c|}
\hline
 \#  & $\Delta b'^k$  & Q  & U & D & L, H & E \\  \hline \hline
  1  &  7   & $-2,-2,-2$ & $-1,-1,-1$ & $1,-1,-1$
& $-2,-2,-2-,2,-3$ & $-1,-1,-1$ \\ \hline
  2  &  6   & $-2,-2,-2$ & $-1,-1,-2$ & $1,-1,-1$
& $-2,-2,-2-,2,-3$ & $-1,-1,-1$ \\ \hline
  3  &  5   & $-1,-2,-2$ & $-1,-1,-2$ & $1,-1,-1$
& $-2,-2,-2-,2,-3$ & $-1,-1,-1$ \\ \hline
  4  &  4   & $-1,-2,-2$ & $-1,-2,-2$ & $1,-1,-1$
& $-2,-2,-2-,2,-3$ & $-1,-1,-1$ \\
\hline
                                   \end{tabular}

\newpage
{\bf \large Table 8-1. Minimal String Models from $Z_3 \times Z_6$ and
$Z_6 \times Z_6$ orbifolds with $\Delta b'\geq 6$}

\begin{tabular}{|c|c|c|c|c|c|c|}
\hline
 \#  & $\Delta b'^k$  & Q  & U & D & L, H & E \\  \hline \hline
  1  &  8   & $-2,-2,-2$ & $0,-1,-1$ & $1,-1,-1$
& $-2,-2,-2-,2,-3$ & $-2,-2,-2$ \\ \hline
  2  &  7   & $-1,-2,-2$ & $0,-1,-1$ & $1,-1,-1$
& $-2,-2,-2-,2,-3$ & $-2,-2,-2$ \\
  3  &  7   & $-2,-2,-2$ & $0,-1,-2$ & $0,-1,-1$
& $-2,-2,-2-,2,-3$ & $-2,-2,-2$ \\
  4  &  7   & $-2,-2,-2$ & $0,-1,-1$ & $1,-1,-2$
& $-2,-2,-2-,2,-3$ & $-2,-2,-2$ \\
  5  &  7   & $-2,-2,-2$ & $-1,-1,-1$ & $1,-1,-1$
& $-2,-2,-2-,2,-3$ & $-1,-1,-1$ \\
  6  &  7   & $-2,-2,-2$ & $-1,-1,-1$ & $1,-1,-1$
& $-2,-2,-2-,2,-3$ & $-1,-2,-2$ \\
  7  &  7   & $-2,-2,-2$ & $-1,-1,-1$ & $1,-1,-1$
& $-2,-2,-2-,2,-3$ & $-2,-2,-2$ \\
  8  &  7   & $-2,-2,-2$ & $0,-1,-2$ & $1,-1,-1$
& $-2,-2,-2-,2,-3$ & $-1,-1,-1$ \\
  9  &  7   & $-2,-2,-2$ & $0,-1,-2$ & $1,-1,-1$
& $-2,-2,-2-,2,-3$ & $-1,-2,-2$ \\
 10  &  7   & $-2,-2,-2$ & $0,-1,-2$ & $1,-1,-1$
& $-2,-2,-2-,2,-3$ & $-2,-2,-2$ \\ \hline
 11  &  6   & $-1,-1,-2$ & $0,-1,-1$ & $1,-1,-1$
& $-2,-2,-2-,2,-3$ & $-2,-2,-2$ \\
 12  &  6   & $-1,-2,-2$ & $0,-1,-1$ & $0,-1,-1$
& $-2,-2,-2-,2,-3$ & $-2,-2,-2$ \\
 13  &  6   & $-1,-2,-2$ & $0,-1,-1$ & $1,-1,-2$
& $-2,-2,-2-,2,-3$ & $-2,-2,-2$ \\
 14  &  6   & $-1,-2,-2$ & $-1,-1,-1$ & $1,-1,-1$
& $-2,-2,-2-,2,-3$ & $-1,-2,-2$ \\
 15  &  6   & $-1,-2,-2$ & $-1,-1,-1$ & $1,-1,-1$
& $-2,-2,-2-,2,-3$ & $-2,-2,-2$ \\
 16  &  6   & $-1,-2,-2$ & $0,-1,-2$ & $1,-1,-1$
& $-2,-2,-2-,2,-3$ & $-1,-2,-2$ \\
 17  &  6   & $-1,-2,-2$ & $ 0,-1,-2$ & $1,-1,-1$
& $-2,-2,-2-,2,-3$ & $-2,-2,-2$ \\
 18  &  6   & $ 0,-2,-2$ & $ 0,-1,-1$ & $1,-1,-1$
& $-2,-2,-2-,2,-3$ & $-2,-2,-2$ \\
 19  &  6   & $-2,-2,-2$ & $-1,-1,-1$ & $1,-1,-1$
& $-1,-2,-2-,2,-3$ & $-2,-2,-2$ \\
 20  &  6   & $-2,-2,-2$ & $-1,-1,-1$ & $1,-1,-1$
& $-2,-2,-2-,2,-2$ & $-2,-2,-2$ \\
 21  &  6   & $-2,-2,-2$ & $-1,-1,-1$ & $0,-1,-1$
& $-2,-2,-2-,2,-3$ & $-2,-2,-2$ \\
 22  &  6   & $-2,-2,-2$ & $-1,-1,-1$ & $1,-1,-2$
& $-2,-2,-2-,2,-3$ & $-2,-2,-2$ \\
 23  &  6   & $-2,-2,-2$ & $-1,-1,-2$ & $1,-1,-1$
& $-2,-2,-2-,2,-3$ & $-1,-1,-1$ \\
 24  &  6   & $-2,-2,-2$ & $-1,-1,-2$ & $1,-1,-1$
& $-2,-2,-2-,2,-3$ & $-1,-2,-2$ \\
 25  &  6   & $-2,-2,-2$ & $-1,-1,-2$ & $1,-1,-1$
& $-2,-2,-2-,2,-3$ & $-2,-2,-2$ \\
 26  &  6   & $-2,-2,-2$ & $ 0,-1,-2$ & $1,-1,-1$
& $-1,-2,-2-,2,-3$ & $-2,-2,-2$ \\
 27  &  6   & $-2,-2,-2$ & $ 0,-1,-2$ & $1,-1,-1$
& $-2,-2,-2-,2,-2$ & $-2,-2,-2$ \\
 28  &  6   & $-2,-2,-2$ & $ 0,-1,-2$ & $0,-1,-1$
& $-2,-2,-2-,2,-3$ & $-2,-2,-2$ \\
 29  &  6   & $-2,-2,-2$ & $ 0,-1,-2$ & $1,-1,-2$
& $-2,-2,-2-,2,-3$ & $-2,-2,-2$ \\
 30  &  6   & $-2,-2,-2$ & $ 0,-2,-2$ & $1,-1,-1$
& $-2,-2,-2-,2,-3$ & $-1,-1,-1$ \\
 31  &  6   & $-2,-2,-2$ & $ 0,-2,-2$ & $1,-1,-1$
& $-2,-2,-2-,2,-3$ & $-1,-2,-2$ \\
 32  &  6   & $-2,-2,-2$ & $ 0,-2,-2$ & $1,-1,-1$
& $-2,-2,-2-,2,-3$ & $-2,-2,-2$ \\
\hline
                                   \end{tabular}
\newpage
{\bf \large Table 8-2. Minimal String Models from $Z_3 \times Z_6$ and
$Z_6 \times Z_6$ orbifolds with $\Delta b'=5$}
\vspace{5mm}

\begin{tabular}{|c|c|c|c|c|c|c|}
\hline
 \#  & $\Delta b'^k$  & Q  & U & D & L, H & E \\  \hline \hline
  1  &  5   & $-1,-2,-2$ & $-1,-1,-2$ & $1,-1,-1$
& $-2,-2,-2-,2,-3$ & $-1,-1,-1$ \\
  2  &  5   & $-1,-2,-2$ & $0,-2,-2$ & $1,-1,-1$
& $-2,-2,-2-,2,-3$ & $-1,-1,-1$ \\
  3  &  5   & $ 0,-2,-2$ & $0,-1,-1$ & $0,-1,-1$
& $-2,-2,-2-,2,-3$ & $-2,-2,-2$ \\
  4  &  5   & $ 0,-2,-2$ & $0,-1,-1$ & $1,-1,-2$
& $-2,-2,-2-,2,-3$ & $-2,-2,-2$ \\
  5  &  5   & $ 0,-2,-2$ & $-1,-1,-1$ & $1,-1,-1$
& $-2,-2,-2-,2,-3$ & $-1,-2,-2$ \\
  6  &  5   & $-2,-2,-2$ & $-1,-1,-1$ & $-1,-1,-1$
& $-2,-2,-2-,2,-3$ & $-2,-2,-2$ \\
  7  &  5   & $-2,-2,-2$ & $-1,-1,-1$ & $0,-1,-2$
& $-2,-2,-2-,2,-3$ & $-2,-2,-2$ \\
  8  &  5   & $-2,-2,-2$ & $-1,-1,-1$ & $1,-2,-2$
& $-2,-2,-2-,2,-3$ & $-2,-2,-2$ \\
  9  &  5   & $-2,-2,-2$ & $-1,-1,-2$ & $0,-1,-1$
& $-2,-2,-2-,2,-3$ & $-1,-2,-2$ \\
 10  &  5   & $-2,-2,-2$ & $-1,-1,-2$ & $1,-1,-2$
& $-2,-2,-2-,2,-3$ & $-1,-2,-2$ \\
 11  &  5   & $-2,-2,-2$ & $0,-1,-2$ & $-1,-1,-1$
& $-2,-2,-2-,2,-3$ & $-2,-2,-2$ \\
 12  &  5   & $-2,-2,-2$ & $0,-1,-2$ & $0,-1,-2$
& $-2,-2,-2-,2,-3$ & $-2,-2,-2$ \\
 13  &  5   & $-2,-2,-2$ & $0,-1,-2$ & $1,-2,-2$
& $-2,-2,-2-,2,-3$ & $-2,-2,-2$ \\
 14  &  5   & $-2,-2,-2$ & $-1,-2,-2$ & $1,-1,-1$
& $-2,-2,-2-,2,-3$ & $-1,-1,-2$ \\
 15  &  5   & $-2,-2,-2$ & $ 0,-2,-2$ & $0,-1,-1$
& $-2,-2,-2-,2,-3$ & $-1,-2,-2$ \\
 16  &  5   & $-2,-2,-2$ & $0,-2,-2$ & $1,-1,-2$
& $-2,-2,-2-,2,-3$ & $-1,-2,-2$ \\ \hline
                                   \end{tabular}

\newpage
{\bf \large Table 8-3. Minimal String Models from $Z_3 \times Z_6$ and
$Z_6 \times Z_6$ orbifolds with $\Delta b'=4$}
\vspace{5mm}

\begin{tabular}{|c|c|c|c|c|c|c|}
\hline
 \#  & $\Delta b'^k$  & Q  & U & D & L, H & E \\  \hline \hline
  1  &  4   & $ 0,-1,-1$ & $ 0,-1,-1$ & $1,-1,-1$
& $-2,-2,-2-,2,-3$ & $-1,-2,-2$ \\
  2  &  4   & $ 0,-1,-2$ & $ 0,-1,-2$ & $1,-1,-1$
& $-2,-2,-2-,2,-3$ & $-1,-2,-2$ \\
  3  &  4   & $-1,-2,-2$ & $-1,-2,-2$ & $1,-1,-1$
& $-2,-2,-2-,2,-3$ & $-1,-1,-1$ \\
  4  &  4   & $-1,-2,-2$ & $-1,-1,-2$ & $1,-1,-1$
& $-2,-2,-2-,2,-3$ & $-1,-1,-2$ \\
  5  &  4   & $-1,-2,-2$ & $-1,-2,-2$ & $1,-1,-1$
& $-2,-2,-2-,2,-3$ & $-1,-2,-2$ \\
  6  &  4   & $ 0,-2,-2$ & $-1,-1,-1$ & $1,-1,-1$
& $-1,-2,-2-,2,-3$ & $-2,-2,-2$ \\
  7  &  4   & $ 0,-2,-2$ & $-1,-1,-1$ & $1,-1,-1$
& $-2,-2,-2-,2,-2$ & $-2,-2,-2$ \\
  8  &  4   & $ 0,-2,-2$ & $ 0,-1,-2$ & $1,-1,-1$
& $-1,-2,-2-,2,-3$ & $-2,-2,-2$ \\
  9  &  4   & $ 0,-2,-2$ & $ 0,-1,-2$ & $1,-1,-1$
& $-2,-2,-2-,2,-2$ & $-2,-2,-2$ \\
 10  &  4   & $ 0,-2,-2$ & $ 0,-2,-2$ & $1,-1,-1$
& $-2,-2,-2-,2,-3$ & $-1,-2,-2$ \\
 11  &  4   & $-2,-2,-2$ & $-1,-1,-2$ & $0,-1,-1$
& $-1,-2,-2-,2,-3$ & $-2,-2,-2$ \\
 12  &  4   & $-2,-2,-2$ & $-1,-1,-2$ & $0,-1,-1$
& $-2,-2,-2-,2,-2$ & $-2,-2,-2$ \\
 13  &  4   & $-2,-2,-2$ & $-1,-1,-2$ & $1,-1,-2$
& $-1,-2,-2-,2,-3$ & $-2,-2,-2$ \\
 14  &  4   & $-2,-2,-2$ & $-1,-2,-2$ & $1,-1,-2$
& $-2,-2,-2-,2,-2$ & $-2,-2,-2$ \\
 15  &  4   & $-2,-2,-2$ & $-1,-2,-2$ & $1,-1,-1$
& $-2,-2,-2-,2,-2$ & $-1,-2,-2$ \\
 16  &  4   & $-2,-2,-2$ & $-1,-2,-2$ & $0,-1,-1$
& $-2,-2,-2-,2,-3$ & $-1,-1,-1$ \\
 17  &  4   & $-2,-2,-2$ & $-1,-2,-2$ & $1,-1,-2$
& $-2,-2,-2-,2,-3$ & $-1,-1,-1$ \\
 18  &  4   & $-2,-2,-2$ & $ 0,-2,-2$ & $0,-1,-1$
& $-1,-2,-2-,2,-3$ & $-2,-2,-2$ \\
 19  &  4   & $-2,-2,-2$ & $ 0,-2,-2$ & $0,-1,-1$
& $-2,-2,-2-,2,-2$ & $-2,-2,-2$ \\
 20  &  4   & $-2,-2,-2$ & $ 0,-2,-2$ & $0,-1,-2$
& $-1,-2,-2-,2,-3$ & $-2,-2,-2$ \\
 21  &  4   & $-2,-2,-2$ & $ 0,-2,-2$ & $1,-1,-2$
& $-1,-2,-2-,2,-3$ & $-2,-2,-2$ \\
\hline
                                   \end{tabular}

\end{document}